From kirk@kirk0.mpi-hd.mpg.de Mon Jan 22 15:20:12 1996
Received: by kirk0.mpi-hd.mpg.de (5.65/Ultrix3.0-C)
	id AA12407; Mon, 22 Jan 1996 15:20:07 +0100
Date: Mon, 22 Jan 1996 15:20:07 +0100
From: kirk@kirk0.mpi-hd.mpg.de (J. Kirk)
Message-Id: <9601221420.AA12407@kirk0.mpi-hd.mpg.de>
To: masti@boris
Subject: final version - TeX
Status: RO

\documentstyle[epsf]{l-aa}
\newcommand{\gdot}{\stackrel{.}{\gamma}}
\newcommand{\eave}{\epsilon_{\rm av}}
\newcommand{\ledd}{\ell_{{\rm Edd}}}

\newcommand{\gammaeq}{\gamma_{\rm eq}}
\newcommand{\rin}{r_{\rm in}}

\newcommand{\Tin}{T_{\rm in}}

\newcommand{\rgrav}{r_{\rm g}}
\newcommand{\msolar}{M_\odot}

\newcommand{\eqb}{\begin{eqnarray}}
\newcommand{\eqe}{\end{eqnarray}}
\newcommand{\gesim}{\,\raisebox{-0.4ex}{$\stackrel{>}{\scriptstyle\sim}$}\,}
\newcommand{\lesim}{\,\raisebox{-0.4ex}{$\stackrel{<}{\scriptstyle\sim}$}\,}
\begin{document}
\thesaurus{02.01.1; 02.01.2; 11.01.2; 11.02.2 Mkn~421; 11.02.2 Mkn~501; 13.07.3}
\title{On the production of very high energy beamed gamma-rays in 
blazars}
\author{W. Bednarek\inst{1,}\inst{2} \and 
J.G. Kirk\inst{1} \and 
A. Mastichiadis\inst{1}}
\institute{Max-Planck-Institut f\"ur Kernphysik,
Postfach 10 39 80, D-69029 Heidelberg,
Germany
\and
University of \L \'od\'z, Department of Experimental Physics,
ul. Pomorska 149/153, 90-236 \L\'od\'z, Poland}
\offprints{J.G. Kirk}
\date{Received; accepted \dots}
\maketitle
\begin{abstract}
The variable flux of TeV gamma-rays detected from Mkn~421 and Mkn~501 requires 
the presence of high energy electrons, which could in principle produce
large numbers of electron/positron pairs, leading to an electromagnetic
cascade. We point out that this scenario can be avoided if electrons
are accelerated to high energy rectilinearly, rather than being injected isotropically
into a blob, as in most of the models of the GeV gamma-ray emission.
By balancing linear acceleration by an electric field against inverse
Compton losses in the radiation field of the accretion
disk we calculate the emitted spectra and
find the conditions which must be fulfilled in order to
exclude the development of
electromagnetic cascades during acceleration.
Assuming these to be fulfilled, 
we show that the maximum possible photon energy is 
approximately $10M_8^{2/5}\,$TeV, where $M_8$ is the 
mass of the central black hole in units of $10^8\,\msolar$.
In addition we compute the optical depth to absorption 
of TeV photons on a possible
isotropic scattered component and
on the observed nonthermal radiation (in the case of Mkn~421)
and find that TeV photons can escape provided the 
nonthermal X-rays
originate in a jet moving with a Lorentz factor  
$\gamma_{\rm b}\gesim8$.   
\end{abstract}

\section{Introduction}
Gamma-rays of energy up to 10~GeV have been observed from many active 
galactic nuclei by the EGRET experiment on the Compton Gamma-Ray 
Observatory (von Montigny et al.~\cite{vonmontignyetal95}).
Although hadronic models have been proposed (Mannheim \& Biermann~\cite{mannheimbiermann92})
most models assume the gamma-rays are produced by inverse Compton scattering off electrons
contained in a blob of plasma which itself is in relativistic 
motion in the general direction of the observer
(Maraschi et al~\cite{maraschietal92}; Dermer \& 
Schlickeiser~\cite{dermerschlickeiser93}; 
Blandford \& Levinson~\cite{blandfordlevinson95}). 
These models differ in their assumptions about the soft target photons
(Schlickeiser~\cite{schlickeiser96}),
but they all assume
the electrons to be isotropically distributed in the rest frame of the blob.
The gamma-ray luminosity they predict is
relativistically boosted, so that the observer receives a flux which exceeds 
that measured in the rest frame of the blob by several
orders of magnitude.

In the TeV regime two blazars
-- Mkn~421 and Mkn~501 -- 
have been observed to emit gamma-rays of 
energy $\gesim 4\,$TeV (Punch et al.~\cite{punchetal92}, Quinn et al.~\cite{quinnetal95}).
Of those, Mkn~421 is also an EGRET source and its spectrum
is consistent with an unbroken power law spectrum of photon index 
$\simeq -$2 between GeV and TeV energies, which 
makes it unlikely that these photons have been reprocessed by an electromagnetic 
cascade. 
Electrons of Lorentz factor exceeding $10^6$ are required in the blob frame 
in order to explain the observations, and, if isotropically distributed,
these would inevitably undergo photon scattering events well into the Klein-Nishina
regime. Such events produce pairs, either directly via the triplet pair production
process (Mastichiadis~\cite{mastichiadis91}) or indirectly via photon-photon pair production.
Thus, although there are no detailed calculations of the development of a cascade in the 
anisotropic radiation field seen by a relativistic blob, a problem is quite likely to 
arise.

In view of this, we present here a model 
in which electrons are accelerated to 
high energies rectilinearly in an electric field. For
certain values of the potential, acceleration
can be balanced by
inverse Compton scattering on the accretion disk photons  
which occurs in the Thomson regime, thus avoiding the problem of
a cascade.
Since we are interested in the outgoing gamma radiation, 
we follow electrons which propagate outwards and
calculate their energy as well as 
that of the associated photons as a function of height above the accretion disk.
Adopting a simple model for the disk radiation we find that
close to the disk, Thomson losses limit the particle energy to a relatively 
small value, whereas far
from it the losses are smaller and particles reach higher energies before their 
acceleration is saturated. Lower energy gamma rays thus originate
close to the disk whereas photons of  TeV energy come from 
a distance of about 
1000~gravitational radii away from the black hole.

The system of a jet emerging along the rotation
axis of an accretion disk which underlies our calculations is described 
in Sect.~\ref{diskjet}. The saturation of electron acceleration by inverse Compton
scattering and the resulting spectrum is computed in Sect.~\ref{icsaturation}.
In the same Section we give also the necessary conditions which make the
inverse Compton scattering to occur in the Thomson regime. 
Finally we discuss the implications of our model in Sect.~\ref{discussion}

\section{The disk/jet geometry}
\label{diskjet}
We adopt a simple one-dimensional model of a jet in which there are 
localised regions where particles can be accelerated by an electric field.
There are
several ways of realising such a configuration. Magnetic reconnection 
(e.g., Schindler et al~\cite{schindleretal91}, Haswell 
et al~\cite{haswelletal92}) leads to potential drops aligned along the 
magnetic field, but is amenable only to rough estimates. Almost rectilinear 
acceleration can occur for some trajectories as a result of an encounter 
with a relativistic oblique shock 
front, such as may be expected within a relativistic jet 
(Begelman \& Kirk~\cite{begelmankirk90}) and is an integral part
of the \lq surfatron\rq\ process (Katsouleas \& Dawson~\cite{katsouleasdawson83})
which has as yet not been applied in an astrophysical context.
Alternatively, an electric field component along the jet axis may be
induced close to the black hole (Blandford \& 
Znajek \cite{blandfordznajek77}, 
Macdonald \& Thorne \cite{macdonaldthorne82}, Bednarek \& Kirk~\cite{bednarekkirk95}) 
or arise in the sheath between a relativistic jet and its surroundings
(Bisnovaty-Kogan \& Lovelace~\cite{bisnovatykoganlovelace95}).

The orientation of accelerating regions is important for the computation of the 
losses suffered by the particles. We will assume it to be along the jet axis.
The maximum potential which can be realised within an acceleration region
scales with the local magnetic field strength and the coherence length scale 
of the region. For a constant opening angle of the jet, the dominant 
magnetic field component falls off inversely as the distance 
$z\rgrav$ from the black hole. Here $\rgrav$ is the gravitational radius
$\rgrav=2GM_{\rm BH}/c^2$ of a black hole of mass $M_{\rm BH}$. In this case,
the coherence length is likely to be proportional to $z$, so that the potential
available in an acceleration zone is independent of position in the jet
(Bednarek et al~\cite{bednareketal95}). In this paper, we treat a more
general situation in which the maximum potential $\Delta V$ varies inversely
as some power $\alpha$ of $z$, i.e., $\Delta V = V_0 z^{-\alpha}$. The 
coherence length will be taken to be a fixed fraction $\beta$ ($<1$) 
of the distance from the black hole, so that the electric field is 
given by $E(z)=V_0 z^{-\alpha-1}/(\rgrav\beta)$. 
Although it is to be expected that
the jet moves relativistically with bulk Lorentz factor $\sim10$  
we assume the acceleration
zones to be stationary in the \lq lab.\rq\ frame, as would be expected if they
are associated with shock fronts caused by an obstacle at the edge of the jet.
The particle dynamics are then described simply in terms of the electric field,
 and have no superimposed bulk motion until they have emerged from an 
acceleration zone and have had enough time to isotropise.    

The radiation fields which may be present and lead to inverse Compton losses
 are of two different kinds - an ambient almost isotropic field arising from 
backscattering of the luminosity of the entire black hole/disk/jet system
(Sikora et al~\cite{sikoraetal94}, Blandford 
\& Levinson~\cite{blandfordlevinson95}) and an anisotropic component of photons
which come directly from the disk. Objects which emit TeV photons must have 
relatively weak isotropic radiation fields, so that in this paper we will 
concentrate on the direct radiation from the disk (see Sect.~\ref{discussion}).
An important property of an accretion disk is that its surface temperature
falls off with increasing radius $r$. To include this effect we adopt a
Shakura/Sunyaev profile and assume the disk emits blackbody radiation at 
temperature $T(r)=\Tin (\rin/r)^{3/4}$ where $\Tin$ is the temperature
at the inner edge of the disk assumed to lie at $r=\rin=3\rgrav$. 
Thus, two parameters are required to specify the disk, and these we 
choose to be the luminosity $\ledd$ 
of the disk in units of the Eddington luminosity:  
$\ell_{\rm Edd}=1.5\times10^{-35}\Tin^4\rgrav$ and 
the mass $M_8$ of the black hole, expressed in units of $10^8\,\msolar$.

%
%
%
\section{Inverse Compton scattering}
\label{icsaturation}

The Lorentz factor of electrons accelerated rectilinearly by the electric
field described in Sect.~2, is limited under certain conditions by 
Compton scattering. 
The radiation responsible for this can in principle either come
directly from the disk, or have been scattered by surrounding material
into a quasi-isotropic distribution. 
The Thomson losses of a particle moving
along the jet axis in the anisotropic radiation field of the disk
are straightforward to compute e.g., by generalising the results
of Protheroe et al~(\cite{protheroeetal92}):
\eqb
\gdot=0.25\times\gamma^2\ell_{\rm Edd} M^{-1}_8 z^{-3}\ {\rm{s}}^{-1}
\label{gloss}
\enspace.
\eqe
Using this formula, one finds that losses balance gains in the electric field 
at an equilibrium
Lorentz factor given by 
\eqb
\gammaeq(z)
&=&8.82\times 10^{-5} 
V_0^{1/2}z^{-\alpha/2+1}\beta^{-1/2}\ledd^{-1/2}
\label{gameq}
\eqe 
where $V_0$ is measured in volts.
Losses on the isotropic radiation field depend on
the radius typical of the scattering material $R_{\rm sc}$ and on
its optical depth $\tau_{\rm sc}$.
We find that disk radiation dominates the energy losses
for distances
\eqb
z&\lesim&
8000 (R_{\rm sc}/1\,{\rm pc})^{2/3}
(\tau_{\rm sc}/10^{-3})^{-1/3}
 M_8^{-2/3} 
\enspace.
\label{isotrop}
\eqe
A limit on the column density of matter in Mkn~421 can be
obtained from X-ray 
observations: $n_{\rm H}\le 1.45\times 10^{20}$cm$^{-2}$ 
(Fink et al~\cite{finketal91}) corresponding to a Thomson optical depth 
$\tau_{\rm sc}\le 10^{-4}$. According to Eq.~(\ref{isotrop}) this
means that losses are dominated by the radiation field 
of the disk.

Equations~(\ref{gloss}) and (\ref{gameq}) are valid 
provided the electrons scatter in the Thomson regime.
Approximating the blackbody spectrum by monoenergetic photons
of energy $3k_{\rm B}T$, this condition implies the requirement
\eqb
3k_{\rm B}\Tin z^{-3/4}\left\lbrace
(3z/r)^{3/4}\left[1-{({1+(z/r)^2})^{-1/2}}\right]\right\rbrace\gamma
\lesim mc^2
\enspace,
\eqe
for photons from a point at radius $r$ on the disk.
The quantity in braces on the left-hand side 
reaches a maximum of roughly 0.7 at $r\approx2z$, leading to 
\eqb
\gamma &\lesim&
1.4\times10^4 (M_8/\ledd)^{1/4} z^{3/4}
\enspace.
\eqe

In order
to channel energy efficiently into gamma-rays, the potential must accelerate
electrons to an energy high enough for losses to be important. If this is not
the case, the acceleration process puts energy primarily into particles,
which may subsequently isotropise and radiate, as in other models. 
Thus, the condition $\gammaeq mc^2 = e \Delta V$ separates the region
in which energy is put directly into radiation, which we call the 
\lq radiation dominated\rq\ zone, from the \lq particle
dominated zone\rq\ where energy is first transferred by the acceleration 
mechanism into particles, which cool only after acceleration has ceased. 

Figure~1 shows the constraints on the parameter space for black-hole mass 
$M_8=1$, disk luminosity $\ledd=0.1$ and $\beta=.1$. 
In addition, the dependence of $\gammaeq$ on $z$ is shown for
various values of the potential $V_0$ and for $\alpha=0$. 
\begin{figure}[t]
\epsfxsize=8 cm
\epsffile{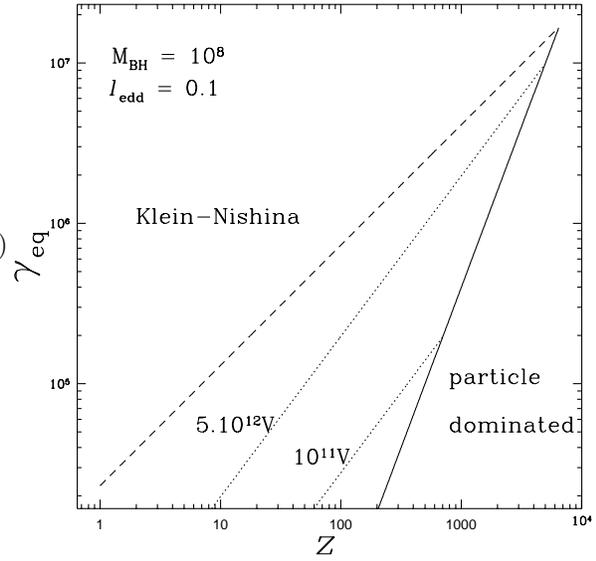}
\caption{
The parameter space for rectilinear acceleration. The solid line shows the 
constraint
$\protect\gammaeq=e\protect\Delta V/(mc^2)$ dividing radiation and particle
dominated zones. The dashed line shows the boundary between the Thomson and 
Klein-Nishina scattering zones. These boundaries are independent of 
the dependence of the potential along the jet and are plotted for 
$\protect\beta=0.1$. The dotted lines show $\protect\gammaeq$ 
for $V_0=10^{11}$ and $V_0=5\protect\times10^{12}\protect\,$V. These lines 
have the slope $1-\protect\alpha/2$ and 
are shown for $\protect\Delta V=\protect\,$constant, corresponding to 
$\protect\alpha=0$.}
\vspace{-0.1 cm}
\end{figure}

If a particle is accelerated linearly at a point in the jet within the 
permitted region lying between the dashed and solid lines in Fig.~1, 
its equilibrium Lorentz factor is given by 
Eq.~(\ref{gameq}). If, however, the electric field is too strong,
and the particle finds itself above the Klein-Nishina boundary, the 
photons it produces will start a pair avalanche similar to that described
by Bednarek \& Kirk~(\cite{bednarekkirk95}). As a result, we can expect that
the particle will not achieve the Lorentz factor given by Eq.~(\ref{gameq}).
On the other hand, a weak electric field places the particle in the 
particle dominated region. Here the total available potential in an
acceleration region is insufficient to produce particles of Lorentz
factor given by Eq.~(\ref{gameq}). Thus, the maximum possible Lorentz factor
$\gamma_{\rm max}$
is achieved for jet parameters within the permitted region and at the highest
$z$ compatible with this restriction.
This value depends only weakly on 
the mass of the black hole
$\gamma_{\rm max}=1.1\times10^{8}M_8^{2/5}\ledd^{1/5}\,\beta^{3/5}$
and is achieved at a height
$z_{\rm max}=1.6\times10^{5}M_8^{1/5}\ledd^{3/5}\,\beta^{4/5}$.

An analytic expression can be found for the 
radiated photon spectra in the \lq\lq radiation dominated\rq\rq\ zone 
using the method of Protheroe et al~(\cite{protheroeetal92}). The 
average energy $\eave$ of a photon produced at height $z$ is approximately
$\gammaeq^2\epsilon_{\rm s}$, where $\epsilon_{\rm s}$ is the energy of 
the dominant soft photons, which are those originating at a point on the 
disk approximately given by $r=\rgrav z$. Thus, 
$\epsilon_{\rm s}\propto z^{-3/4}$ and $\eave\propto z^{-\alpha+5/4}$.
The energy converted into gamma-rays by a single particle passing through
a distance ${\rm d}z$ is equal to the energy extracted from the electric field.
If ${\rm d}n$ photons are thereby produced, one can write 
$\eave{\rm d}n = ec\,E(z){\rm d}z\propto z^{-\alpha-1}{\rm d}z$. 
As shown by Protheroe et al~(\cite{protheroeetal92}) 
the photon spectrum ${\rm d}n/{\rm d}\epsilon$ 
can be found by convolving  ${\rm d}n/{\rm d}\eave$ with
the real distribution $p(\epsilon,\eave)$
of emitted photons. For the cases which are of 
interest here, however, this
does not affect the spectral slope, which is given by
\eqb
({{\rm d}n/{\rm d}\epsilon})&\propto& \epsilon^{-(4\alpha-10)/(4\alpha-5)}
\label{spectrum}
\enspace.
\eqe
If we assume equal numbers of particles are present in the acceleration zones
over a range of heights $z_1<z<z_2$, then we obtain a power-law
spectrum between the photon energies $\epsilon_1=\epsilon(z_1)$ 
and $\epsilon_2=\epsilon(z_2)$ where
\eqb
\epsilon(z)&\simeq&1.9\times10^{-7}V_0\,\ledd^{-3/4}M_8^{-1/4}\beta^{-1}z^{-\alpha+5/4}\ {\rm eV}
\enspace.
\eqe
For $\alpha=0$, a spectrum of index $-2$ is obtained.
The maximum possible photon energy is then
\eqb
\epsilon_{\rm max}&\simeq&3.4\times10^{13}\beta^{3/5}\ledd^{1/5}M_8^{2/5}~{\rm eV}
\label{maxphoton}
\enspace.
\eqe
\section{Discussion}
\label{discussion}  

In our model, the gamma radiation observed from blazars
is produced by electrons subject to rectilinear
acceleration which is 
balanced by losses due to inverse Compton scattering on accretion disk photons.
In this case, the TeV $\gamma$-ray emission from 
an object such as Mkn~421 originates in
the upper part of the radiation dominated zone of the jet at 
a distance of about 1000 gravitational radii from the black hole. 
If the disk radiation dominates over that of the isotropic
component at even larger distances ($z\approx6000$), 
the maximum achievable photon energy for
$\beta\approx\ledd\approx0.1$ is about $10\,$TeV.

There are two interesting features of the proposed model. 
Firstly, the inclusion of an explicit acceleration mechanism
leads naturally to a prediction of the site at which 
high energy gamma rays are produced.
Secondly, the fact that
the electron-soft photon collisions occur in the Thomson regime guarantees
that the resulting gamma rays are emitted below the threshold for pair production, thus
avoiding the complications of an electromagnetic cascade.

However, severe additional constraints 
must be fulfilled if these photons are to leave the source
and avoid absorption on other ambient radiation fields. 
Two types of field are important: 
(i) an isotropic component arising from scattered disk 
radiation and (ii) the observed nonthermal radiation. Of these,
(i) is not directly observed in the blazars of relevance here.
Thus, there is little information 
concerning how much matter 
surrounds the accretion disk 
(such as column density or typical
distance scale). 
Detailed computations of the optical depth
for $\gamma$-ray photons escaping from the accretion disk
in Mkn~421 have been performed by B\"ottcher \& Dermer~(\cite{bottcherdermer95}).
However, these authors assume a Thomson optical depth around the disk 
of 0.02 out to a 
distance of 0.1~pc. This is consistent with ROSAT observations 
(Fink et al~\cite{finketal91}) only if 
the X-rays are produced at $z\rgrav>0.1$~pc.  
If, as we assume, this is not the case,  $\tau_{\rm sc}<10^{-4}$ and the absorption of TeV 
$\gamma$-rays by the isotropic component (i) is negligible.

\begin{figure}[t]
\epsfxsize=8 cm
\epsffile[50 175 385 500]{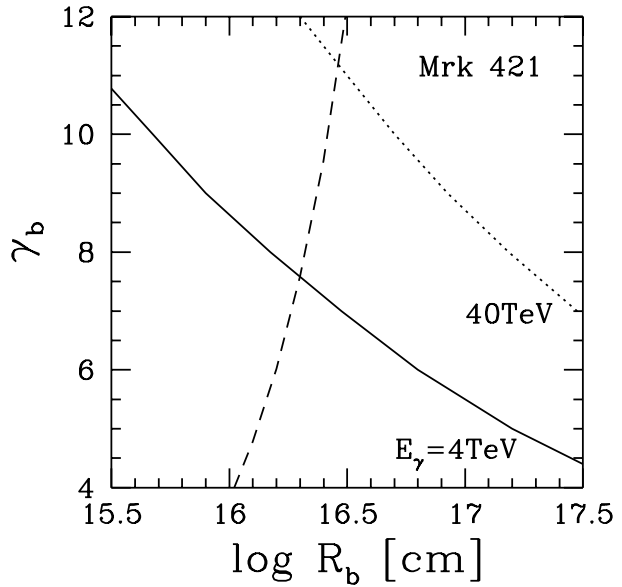}
\caption{The requirements for escape of 4~TeV and
40~TeV $\gamma$-rays 
through the nonthermal radiation of a relativistic
blob in Mkn~421. 
The Lorentz factor $\gamma_{b}$ 
and radius $R_{b}$ of a radiating blob above which the optical
depth for 4~TeV (full line) and 40~TeV (dotted line) $\gamma$-rays is less than unity.
The dashed line shows the minimum $\gamma_{\rm b}$ which permits an
X-ray variability time scale of one day. 
}
\vspace{-0.1 cm}
\end{figure}
In case (ii), the nonthermal radiation is directly observed. According to our
model, these photons are produced in the particle dominated zone 
at $z>1000$ where the electrons have time to isotropise and 
form a relativistic radiating
blob as in other models of the GeV gamma-ray emission.
The optical depth to absorption of the TeV gamma-rays in
such a blob is given by
\eqb
\tau(\epsilon)
&=&
{16 D^2\over\gamma_{\rm b}^2 R_{\rm b}c}
\int
{\rm d}\epsilon_{\rm b}
\int{\rm d}(\cos\theta_{\rm b})\, 
{F}(2\gamma_{\rm b}\epsilon_{\rm b})
\nonumber\\
&\times&
(1-\cos\theta_b) 
\sigma_{\gamma\gamma}[\epsilon/(2\gamma_{\rm b}),\epsilon_{\rm b},\theta_{\rm b}]
\enspace,
\eqe
where $D$ is the distance to Mkn~421, $R_{\rm b}$ is the radius of the 
blob, ${F}(\epsilon)$ is the 
observed photon flux at energy $\epsilon$
[which we approximate as a power law with index $-2$ below 
and $-2.4$ above $10^{-4}$ MeV, with a normalisation
taken from the observations 
(Fink et al~\cite{finketal91}, Makino et al~\cite{makinoetal87})] 
and 
$\sigma_{\gamma\gamma}(\epsilon,\epsilon',\theta)$ 
is the cross section for 
$e^{\pm}$ pair production of a photon of energy $\epsilon$ off one of energy
$\epsilon'$ with $\theta$ the angle between the directions of travel
(Jauch \& Rohrlich~\cite{jauchrohrlich55}).

The full line in Fig.~2 shows the parameters of the blob in Mkn~421 
($\gamma_{\rm b}$, $R_{\rm b}$) for which the optical depth for 4~TeV 
$\gamma$-rays is equal to unity. An additional constraint on the 
Lorentz factor of the jet arises from the   
X-ray variability time scale $t_{\rm var}\approx 1$ day 
(e.g. Fink et al~\cite{finketal91}). Assuming that the blob moves directly 
towards the observer we find 
\eqb
\gamma_{\rm b}&>& R_{\rm b}/ct_{\rm var}\,>\,3.8\times10^{-16}R_{\rm b}\ [{\rm cm}]
\enspace,
\eqe
which is shown in Fig.~2 by the dashed line. The region in Fig.~2 
above the full and dashed lines gives the permitted parameters of the blob in 
Mkn~421 for which the $\gamma$-rays with 
maximal energy of 4~TeV can escape. From it we may also 
constrain the Lorentz factor of the blob:
$\gamma_{\rm b}>7.5$.
For reasonable values of $\gamma_{\rm b}$ (of the order of tens) and
an opening angle of the jet of the order of a few degrees (which 
determines $R_{\rm b}$), the blob
is located inside the particle dominated zone of the jet, in 
agreement with our assumptions.

\acknowledgements{W.B. thanks Max Planck Society for the grant of a visiting 
fellowship. This work was partially supported by the Deutsche 
Forsch\-ungs\-gemein\-schaft under Sonder\-forsch\-ungs\-be\-reich~328}.

{}

\begin{thebibliography}{}

\bibitem[1995]{bednarekkirk95}
Bednarek, W., Kirk. J.G. 1995, A\&A 294, 366
%
\bibitem[1995]{bednareketal95}
Bednarek, W., Kirk, J.G., Mastichiadis, A. 1995, Proc.\ 24th ICRC (Rome),
2, 295
%
\bibitem[1990]{begelmankirk90}
Begelman, M.C., Kirk, J.G. 1990, ApJ 353, 66
%
\bibitem[1995]{bisnovatykoganlovelace95}
Bisnovaty-Kogan, G.S., Lovelace, R.V.E. 1995, A\&A 296, L17
%
\bibitem[1995]{blandfordlevinson95}
Blandford, R.D., Levinson, A. 1995, ApJ 441, 79
%
\bibitem[1977]{blandfordznajek77}
Blandford, R.D., Znajek, R. 1977, MNRAS 179, 433
%
\bibitem[1995]{bottcherdermer95}
B\"ottcher, M., Dermer, C.D. 1995, A\&A, 302, 37
%
\bibitem[1993]{dermerschlickeiser93}
Dermer, C.D., Schlickeiser, R. 1993, ApJ 416, 458 (\lq\lq DS\rq\rq)
%
\bibitem[1991]{finketal91}
Fink, H.H., Thomas, H.-C., Hasinger, G. et al. 1991, A\&A 246, L6
%
\bibitem[1992]{haswelletal92}
Haswell, C.A., Tajima, T., Sakai, J. 1992, ApJ 401, 495
%
\bibitem[1955]{jauchrohrlich55}
Jauch, J.M., Rohrlich, F. 1955, The Theory of photons and electrons 
(Addison-Wesley)
%
\bibitem[1983]{katsouleasdawson83}
Katsouleas, T., Dawson, J.M.  1983, Phys.~Rev.~Letts.~51, 392
%
\bibitem[1982]{macdonaldthorne82}
Macdonald, D., Thorne, K.S. 1982, MNRAS 198, 345
%
\bibitem[1987]{makinoetal87}
Makino, F. et al. 1987, ApJ 313, 662
%
\bibitem[1992]{mannheimbiermann92}
Mannheim, K., Biermann, P.L. 1992, A\&A 221, 211
%
\bibitem[1992]{maraschietal92}
Maraschi, L., Ghisellini, G., Celloti, A. 1992, ApJ 397, L5
%
\bibitem[1991]{mastichiadis91}
Mastichiadis, A. 1991, MNRAS 253, 235
%
\bibitem[1992]{protheroeetal92}
Protheroe, R.J., Mastichiadis, A., Dermer, C.D. 1992, Astropart. Phys. 1, 113 
%
\bibitem[1992]{punchetal92}
Punch, M. et al. 1992, Nature 358, 477
%
\bibitem[1995]{quinnetal95}
Quinn, J. 1995, IAU Circ., No. 6169
%
\bibitem[1991]{schindleretal91}
Schindler, K., Hesse, M., Birn, J. 1991, ApJ 380, 293
%
\bibitem[1996]{schlickeiser96}
Schlickeiser, R. 1996, Space Science Reviews, in press
%
\bibitem[1994]{sikoraetal94}
Sikora, M., Begelman, M.C., Rees, M.J. 1994, ApJ 421, 153
%
\bibitem[1995]{vonmontignyetal95}
von Montigny, C., Bertsch, D.L., Chiang, J. et al. 1995, ApJ 440, 525

\end{thebibliography}
\end{document}